\documentclass[11pt]{article}
\usepackage{blindtext}
\usepackage{titlesec}
\usepackage{hyperref}
\usepackage{amsmath,amsfonts,amssymb}
\hypersetup{dvips,dvipdfm,colorlinks=true,urlcolor=magenta,filecolor=magenta,linktoc=page,citecolor=red,linkcolor=blue,bookmarks=true}
\usepackage{array}
\usepackage{graphicx,epstopdf}
\usepackage{multirow}
\usepackage{xcolor}
\title{Raychaudhuri equation and the dynamics of cosmic evolution}
\date{ }
\begin{document}
	\author{Madhukrishna Chakraborty \footnote{chakmadhu1997@gmail.com}~~and~~Subenoy Chakraborty\footnote{schakraborty.math@gmail.com (corresponding author)}
		\\Department of Mathematics, Jadavpur University, Kolkata - 700032, India}
	\maketitle
	\begin{abstract}
		The paper deals with the Raychaudhuri equation (RE) which is a non-linear ordinary differential equation in $\Theta$, the expansion scalar corresponding to a geodesic flow. Focusing theorem which follows as a consequence of the RE has been restated in terms of the cosmic parameter $q$ (deceleration parameter) both for Einstein gravity and for modified gravity theories. Measurable quantities namely the luminosity distance and density parameter are shown to have an upper bound using the Raychaudhuri scalar. An analogy between geometric and cosmological RE has been made. Subsequently,  to find the solution of the non-linear RE a transformation of variable related to the metric scalar of the hyper-surface has been identified which converts the former to a second order differential equation. Finally, the first integral of this second order differential equation gives the entire picture of the dynamics of cosmic evolution.
	\end{abstract}
	\maketitle
	\section{Introduction}
	The standard cosmology has been going through a remarkable phase due to a series of observational precision \cite{SupernovaSearchTeam:1998fmf}-\cite{SDSS:2005xqv} made during the last two decades. According to these observations the universe at present is undergoing an era of accelerated expansion. This phenomenon may be studied by considering congruence of geodesics in a coordinate independent way. A useful tool to analyze the evolution or behavior of geodesics is the Focusing Theorem (FT) \cite{Poisson:2009pwt} which follows as a consequence of an evolution equation for expansion scalar (defined in detail in the subsequent sections) popularly known as the Raychaudhuri equation (RE) named after Amal Kumar Raychaudhuri \cite{Raychaudhuri:1953yv}. RE \cite{raychaudhuri} can be derived by considering the flow of a congruence of geodesic in a deformable medium. The equation is a non-linear differential equation and hence it is not so easy to find out its analytic solution. In this paper we have adopted a technique to find the analytic solution and used this solution to describe the dynamics of cosmic evolution.

	On the other hand, after the detection of gravitational waves \cite{LIGOScientific:2017vwq} Einstein's General Theory of Relativity is the most well accepted theory of gravity to describe physical reality. Assuming the Einstein's equations and using the energy conditions it can be shown that an initially converging congruence of time-like and null geodesics begin to focus (develop a caustic) until zero size within finite value of the affine parameter (for time-like geodesic congruence proper time $\tau$ is the affine parameter). This results in the formation of congruence singularity.  Here comes the notion of ``caustics". There is a connection between ``caustics" and ``singularity". As a result of FT, the congruence will develop a ``caustic" which is nothing but a point at which all geodesics focus. Thus a caustic may also be called as a focal point which is nothing but a singularity of the congruence and not necessarily a singularity of the space-time. This can be understood from two examples--(i) If one considers the null version of RE in the context of gravitational lensing then a caustic or focal point in this case is the intersection of trajectories representing light rays and is known as the caustic of the bundle of trajectories. (ii) The focusing condition given by $R_{\mu\nu}u^{\mu}u^{\nu}\geq0$ is trivially satisfied by flat space-times but it has no singularities. This shows that ``caustic" is not always a singularity of the space-time but essentially a congruence singularity.  However some extra assumptions (discussed later in this section) and the notion of geodesic incompleteness are the additional criteria for the existence of space-time singularity. In this respect, the key turning point was the 1965 singularity theorem by Penrose and a subsequent work of Hawking. They proved the existence of black hole and cosmological singularity using the notion of incompleteness of null and time-like geodesics respectively. The universal attractive nature of gravity in General Relativity is a feature embodied by the RE which requires the non-increasing expansion of a congruence of geodesics.  In general, though no singularity but rather a caustic is formed along the flow lines of the $u^{\mu}$--congruence. This property is usually called the focusing effect on causal geodesics. Although the singularity here is the congruence singularity and may not be a space-time singularity but these conditions along with some global arguments may lead to space-time singularity in certain cases. Although Raychaudhuri pointed out the connection of his equation to the existence of singularities in his 1955 article, however more general results based on global techniques in Lorentzian space-times appeared in the form of singularity theorems following the work of Penrose \cite{Senovilla:2014gza} and Hawking \cite{Hawking:1970zqf}. According to Penrose (for null geodesics) and Hawking (for time-like geodesics) the singularity to be a black hole or cosmological singularity, the incompleteness of geodesic is essential. They have shown that this incompleteness of geodesic is related to the causal structure of the space-time. They proved the existence of singularities by assuming Lorentz signature metrics and causality, the generic conditions on Riemann tensor components, the existence of trapped surfaces and energy conditions on matter. This vital consequence of the RE ultimately proved the inherent existence of singularity in Einstein gravity and played a key role in the proof of the seminal singularity theorems furnished by Hawking and Penrose. In the proof of these theorems singularity was looked upon in terms of geodesic incompleteness. If a space-time manifold has a congruence of closed geodesic which are not complete then the space-time has singularity. The notion of geodesic incompleteness facilitated proof of certain detailed theorems \cite{Hawking:1970zqf}-\cite{Steinbauer:2022hvq}. A space-time singularity marks not only the breakdown of Einstein Gravity in particular but also physics in general---`` End of Everything".\\
	The RE has profound applications in the recent developments in holography principle and quantum computation of black-hole entropy. RE has been explored in various classical and quantum settings. For ref. see \cite{Dadhich:2007pi}-\cite{Dadhich:2005qr}.  The importance of RE lies not only in the singularity analysis in Einstein gravity but also in extended theories of gravity \cite{Chakraborty:2023ork}-\cite{Choudhury:2021zij}. Analog Raychaudhuri equation in mechanics has been studied in \cite{Bhatt:2021hdi}, \cite{Dasgupta:2007nr}. The implication of  RE in anisotropic space-time is presented in \cite{Chakraborty:2023rgb}.  Also some recent approaches have been made to study the quantum RE \cite{Chakraborty:2023voy} and quantum correction to the RE in the context of possible avoidance of singularity \cite{Das:2013oda}. Raychaudhuri equation has been explored in bouncing cosmology as a tool to understand the nature of bouncing point and singularity \cite{Chakraborty:2023lav}. Moreover modified RE becomes an important tool in identifying the black-hole singularity and resolution of it in loop quantum gravity \cite{Blanchette:2020kkk}. In this paper we explore and analyze the Focusing Theorem and how cosmic parameters affect it. Different era of cosmic evolution has been studied in the context of RE and FT. Further a three fold interpretation of the Convergence scalar also known as Raychaudhuri scalar has been given in cosmology which interestingly $\emph converges$ to the same conclusion. Subsequently the non-linear RE has been converted to a second order differential equation in a transformed variable by a suitable transformation related to the metric scalar of the hyper-surface to obtain a first integral from it for a general $n+1$-dimensional space-time. An analogy of this first integral with the first Friedmann equation has been found in Einstein gravity. Finally using the first integral cosmological solution corresponding to each era of cosmic evolution has been explicitly determined in the background of Friedmann–Lemaître–Robertson-Walker (FLRW) space-time model. Since, the solution of the non-linear RE using the first integral method has been applied to understand the dynamics of cosmic evolution therefore this particular treatment adopted in this paper shows an application of non-linear dynamics to cosmology. Further, the paper also illustrates a similar behavior of $\tilde{R}$ and $q$ in the context of convergence via the geometric and cosmological forms of the RE.\\
	The layout of the paper is: Section II deals with the analysis of FT in terms of cosmic parameters and interpretation of the Raychaudhuri scalar in cosmology from three different points of view; In section III, the first order RE has been converted to a second order differential equation by a transformation of variable and a first integral of this second order differential equation has been found. Moreover this section also contains cosmological solutions of different era of evolution of the Universe using this first integral both from analytical and graphical point of view. Finally the paper ends with brief discussion and conclusions of the obtained results in Section IV.
	\section{Focusing theorem in terms of cosmic parameters}
	The general form of the Raychuadhuri equation (RE) for a congruence of time-like geodesics in an ($n+1$)-dimensional space-time  takes the form \cite{Raychaudhuri:1953yv}-\cite{Dasgupta:2007nr},
	\begin{equation}
		\dfrac{d\Theta}{d\tau}=-\dfrac{\Theta^{2}}{n}-2\Sigma^{2}+2\xi^{2}-R_{\mu\nu}u^{\mu}u^{\nu}.\label{eq1}
	\end{equation}
	This equation describes the behavior of time-like geodesics in space-time and is the main ingredient behind the celebrated singularity theorems by Hawking and Penrose. Moreover this equation is purely geometric and is independent of the theory of gravity under consideration. Here the term $\Theta=u^{\mu}_{;\mu}$ is the expansion scalar which describes how geodesics focus or defocus; $2\Sigma^{2}=\sigma^{\mu\nu}\sigma^{\mu\nu}$ is the anisotropy scalar, $\sigma_{\mu\nu}$ is the shear tensor which describes how the shape of the geodesic changes ( for example a circular configuration changes to elliptic one, say); $2\xi^{2}=\omega_{\mu\nu}\omega^{\mu\nu}$ is the rotation scalar, $\omega_{\mu\nu}$ is the vorticity tensor (describes the rotation of the bundle of geodesic); $u^{\mu}$ is the unit tangent vector to the geodesics; $\tau$ is the proper time. The Riemann tensor in mixed form is given by
		\begin{equation}
			R_{ijk}^{s}=\dfrac{\partial \Gamma_{ki}^{s}}{\partial x^{j}}-\dfrac{\partial \Gamma_{ji}^{s}}{\partial x^{k}}+\Gamma_{\sigma j}^{s}\Gamma_{ki}^{\sigma}-\Gamma_{\sigma k}^{s}\Gamma_{ji}^{\sigma}.	
		\end{equation} Here $\Gamma_{ki}^{s}$ is known as Christoffel symbol of second kind and is given by
		\begin{equation}
			\Gamma_{ki}^{s}=\dfrac{1}{2}g^{s\sigma}\left(\dfrac{\partial g_{k\sigma}}{\partial x^{i}}+\dfrac{\partial g_{i\sigma}}{\partial x^{k}}-\dfrac{\partial g_{ki}}{\partial x^{\sigma}}\right)
		\end{equation} where $g_{\mu\nu}$ is the fundamental metric tensor. The Riemann tensor in fully covariant form can be written as 
		\begin{equation}
			R_{a\mu\nu b}=g_{sc}R_{\mu\nu b}^{c}.
		\end{equation}The expressions for Ricci tensor and that of Ricci scalar are given by
		\begin{eqnarray}
			R_{\mu\nu}=g^{ab}R_{a\mu\nu b}\nonumber\\
			R=g^{\mu\nu}R_{\mu\nu}\nonumber.
		\end{eqnarray}This Ricci tensor $R_{\mu\nu}$ is projected along the congruence of geodesics and $\tilde{R}=R_{\mu\nu}u^{\mu}u^{\nu}$ is the Raychaudhuri scalar which plays a pivotal role in focusing of geodesics. The last term on the R.H.S of the Raychaudhuri equation i.e. $-R_{\mu\nu}u^{\mu}u^{\nu}$ highlights the contribution of space-time geometry and it does not depend on the derivative of the vector field. Therefore this term has more general implications than the other terms in (\ref{eq1}). Geometrically, this term can be interpreted as a mean curvature in the direction of $\textbf{u}$ \cite{Albareti:2014dxa}.\\
	We assume the space-time geometry to be homogeneous and isotropic Friedmann–Lemaître–Robertson-Walker (FLRW) model given by 
		\begin{equation}
			ds^{2}=-dt^{2}+a^{2}(t)\left(\dfrac{dr^{2}}{1-\kappa r^{2}}+r^{2}(d\theta^{2}+\sin^{2}\theta d\phi^{2})\right)
		\end{equation}
		where $a(t)$ is the scale factor with cosmic time $t$. $(r,\theta,\phi)$ are co-moving spherical polar coordinates, $\kappa$ is the constant of curvature of the spatial part of the space-time which may have values $0,+1,-1$ for flat, closed and open universe respectively. Conventionally, this metric with signature $(-,+,+,+)$ is used. Thus, the anisotropy scalar $\Sigma=0$. Further, if the time-like geodesics are orthogonal to the space-like hyper-surfaces $t$=constant then by virtue of the Frobenius theorem of differential geometry $\xi=0$. Thus the RE (\ref{eq1}) reduces to much simpler form as 
	\begin{equation}\label{eq2}
		\dfrac{d\Theta}{d\tau}=-\dfrac{\Theta^{2}}{n}-\tilde{R}.
	\end{equation}
	This shows that $\dfrac{d\Theta}{d\tau}+\dfrac{\Theta^{2}}{n}\leq0$ whenever $\tilde{R}\geq0$. Integration of the inequality implies
	\begin{equation}
		\dfrac{1}{\Theta}\geq \dfrac{1}{\Theta_0}+\dfrac{\tau}{n}.\label{eq3}
	\end{equation} Here, $\Theta_{0}$ is the value of $\Theta$ at $\tau=0$ and $\tau\leq0$. Further it is assumed that the geodesic congruence is expanding at $\tau=0$ i.e. $\Theta_{0}>0$ (which refers to the case for an expanding universe). Then from eq. (\ref{eq3}), $\dfrac{1}{\Theta_{0}}=\Theta_{0}^{-1}$ should have passed through zero in some finite time $\tau_{0}$ (say).
	
	$~~~~~~~~~~~~~~~~~~~~~$ In particular $\tau_{0}$ is bounded by the inequality $|\tau_{0}|\leq n\Theta_{0}^{-1}$. This means that at the time $\tau_{0}$, the expansion scalar was infinite i.e. $\Theta(\tau_{0})=-\infty$, which hints that there was a singularity at $\tau_{0}$. Although this only tells that there is a singularity of the geodesic congruence, this analysis is one of the key ingredients for the singularity theorem : If the matter obeys the Strong Energy Condition (SEC) and there exists a positive constant $K>0$ such that $\Theta>K$, everywhere in the past of some specific hyper-surface, then there exists a past singularity where all the past directed geodesics end. This leads to the conclusion that an expanding universe containing matter that obeys the SEC, means that the Universe has a past singularity. This is known as Focusing Theorem (FT) \cite{Albareti:2012se}. Divergence of the expansion parameter ($\Theta$) by itself however does not imply singularity of space-time. But this aided with some global arguments lead to
	spacetime singularity in certain cases. The FT guarantees that if a space-time has a singularity then focusing will occur there. Therefore if one can avoid the focusing, singularity can be removed. Geometrically, singularity can be associated with geodesic incompleteness  as: If a manifold contains congruence of closed geodesics which are complete then there will be no singularity, but for incomplete geodesics there is singularity of the manifold. This notion of geodesic incompleteness ultimately played a key role in the proof
	of the celebrated singularity theorems due to Penrose, Hawking and
	Geroch \cite{Hawking:1970zqf}-\cite{Hawking:1973uf}. 
 A singularity has the property of geodesic incompleteness in which there is a failure to extend either some light path or some particle path beyond a certain affine parameter or proper time. Geodesics are paths of observers through space-time that can only be extended for a finite time as measured by an observer traveling along one. It is generally assumed that at the end of geodesic an observer has fallen into a singularity or has encountered some kind of pathology at which the laws of general relativity break. This geodesic incompleteness leads to the presence of infinite curvatures, making the right hand side of RE infinite. Thus there is a divergence of the expansion parameter $\Theta$. For greater details regarding how the divergence of $\Theta$ is implied by the presence of conjugate point and incomplete geodesic using the geometrical RE is shown in Appendix B for an exact and comprehensive idea to the general readership. Readers may also refer to \cite{Senovilla:2014gza}, \cite{Steinbauer:2022hvq} in this regard.\\

	In Einstein gravity $\tilde{R}=R_{\mu\nu} u^\mu u^\nu=\kappa \left(T_{\mu\nu}u^{\mu}u^{\nu}+\dfrac{1}{2}T\right)$, where $T_{\mu\nu}$ is the energy momentum tensor and $T=g^{\mu\nu}T_{\mu\nu}$. If the matter satisfies the (SEC) i.e. $\tilde{R}\geq0$ then from the RE, FT follows in Einstein Gravity inevitably. Hence RE and hence the FT is a turning point of Einstein gravity that clearly hints the existence of singularity in GR. The positive semi-definiteness of the Raychaudhuri scalar i.e $\tilde{R}\geq0$ is known as the Convergence Condition (CC). Moreover the RE in FLRW background coincides with the acceleration equation (that we obtain from the two Friedmann equations) if we consider the matter to move along hyper-surface orthogonal congruence of time-like geodesics ($\Theta=3\dfrac{\dot{a}}{a}$, $\Sigma=\xi=0$) and this proves the existence of initial big-bang singularity. The RE (\ref{eq1}) is purely a geometric identity and is independent of any theory of gravity. However the effect of gravity comes into picture through the Ricci tensor ($R_{\mu\nu}$). Keeping the same homogeneous or inhomogeneous background RE and FT can be studied in modified gravity theories. Since in extended theories of gravity (other than Einstein gravity but reduce to Einstein gravity in some limits) the field equations are different, it is expected that the underlying models may avoid the cosmological singularity which persists in Einstein gravity under some conditions (for ref. see \cite{Chakraborty:2023ork}, \cite{Choudhury:2021zij})\\
	In Einstein gravity or in usual modified gravity the field equations for gravity can be written as
	\begin{equation}
		G_{\mu\nu}=\kappa~ T_{\mu\nu}^{'}~,
	\end{equation}
	where $T_{\mu\nu}^{'}=T_{\mu\nu}$ is the usual energy-momentum tensor for the matter field in Einstein gravity while $T_{\mu\nu}^{'}=T_{\mu\nu}+T_{\mu\nu}^{(e)}$ for most of the modified gravity theories with $T_{\mu\nu}^{(e)}$ containing the extra geometric/physical terms in the field equations. Thus the Raychaudhuri scalar $\tilde{R}$ takes the following form in terms of the energy-momentum tensor or/and the effective energy-momentum tensor in Einstein gravity (EG) and in Modified gravity (MG) as:\begin{equation}
		\tilde{R} =\kappa~(T_{\mu\nu}-\frac{1}{2}Tg_{\mu\nu})u^{\mu} u^{\nu}=\dfrac{1}{2}\left(\rho+3p\right),~ EG\label{eq4*}\end{equation}
	and,
	\begin{eqnarray}
		\tilde{R} =\kappa~\left[(T_{\mu\nu}-\frac{1}{2}Tg_{\mu\nu})u^{\mu} u^{\nu}+(T^{(e)}_{\mu\nu}-\frac{1}{2}T^{(e)}g_{\mu\nu})u^{\mu} u^{\nu}\right] =\dfrac{1}{2}\left(\rho+3p\right)+\dfrac{1}{2}\left(\rho^{(e)}+3p^{(e)}\right), ~MG.\label{eq5*}
	\end{eqnarray}
	\textbf{\underline{A three fold interpretation of $\tilde{R}$ in cosmology is given below as}}:
	\begin{enumerate}
		\item  In FLRW space-time the (effective) Einstein field equations are
		\begin{eqnarray}
			3H^{2}= \rho,~~2\dot{H}=-(\rho+p) \label{eq4}\\
			3H^{2}=(\rho+\rho_{e}),~~2\dot{H}=-[(\rho+p)+(\rho_{e}+p_{e})]\label{eq5}
		\end{eqnarray} where equation (\ref{eq4}) is for Einstein gravity and (\ref{eq5}) is for modified gravity.
		So the deceleration parameter $
		q=-\left(1+\dfrac{\dot{H}}{H^{2}}\right)$ takes the form (\ref{eq8}) in Einstein gravity and (\ref{eq9}) in Modified gravity as follows :\begin{eqnarray}
			q=\dfrac{\rho+3p}{2\rho}\label{eq8}\\
			~~~~~~~~~~~~~~~~~~~~q=\dfrac{(\rho+3p)+(\rho_{e}+3p_{e})}{2(\rho+\rho_{e})}\label{eq9}\end{eqnarray}
		Hence 
		\begin{eqnarray}
			\tilde{R}=q\rho~, Einstein~gravity\nonumber
			\\ 
			\tilde{R}=q(\rho+\rho_{e})~, Modified~gravity\nonumber
		\end{eqnarray}Thus,
		\begin{equation}\tilde{R}=3qH^{2}\label{eq11}\end{equation}
		for both the cases. Now for convergence $\tilde{R}>0$ so one may conclude that convergence will occur during the evolution of the universe if $q>0$ i.e, CC occurs only in decelerating phase. In other words, $q$ behaves as convergence scalar. This formulation brings out an inherent feature of the deceleration parameter as convergence scalar. This shows that formation of singularity is not possible both in the early inflationary era and in the present late time accelerated era of evolution, while the matter dominated era of evolution favors convergence.
		\item In Einstein gravity expression of $\tilde{R}$ is given by equation (\ref{eq4*}). Let us write,
		\begin{equation}
			\rho=\rho_{1}+\rho_{2},~~p=p_{1}+p_{2}.
		\end{equation}
		Then $\tilde{R}=\dfrac{1}{2}\{\left(\rho_{1}+3p_{1}\right)+\left(\rho_{2}+3p_{2}\right)\}$.  If we assume $(\rho_{1},p_{1})$, the energy density and pressure of normal matter component (that satisfies SEC i.e. $\rho_{1}+3p_{1}\geq0$) then in order to prevent convergence (focusing) we must have
		\begin{equation}
			\rho_{2}+3p_{2}<0, ~~|\rho_{2}+3p_{2}|> \rho_{1}+3p_{1}..
		\end{equation}
		This shows that the component having energy density and pressure $(\rho_{2},p_{2})$ is dark energy. Hence dominance of dark energy over normal matter ( having density and pressure ($\rho_{1},~p_{1}$)) may prevent focusing. Therefore the era dominated by dark energy namely the inflationary era and the present accelerated era of expansion are against the formation of singularity. Further if $\rho_{1}+3p_{1}\geq0$ and $|\rho_{1}+3p_{1}|\geq\rho_{2}+3p_{2}$ then $\tilde{R}\geq0$. Thus the matter dominated era is in favor of convergence. This matches with the conclusion of the previous case.
		\item  Now, we consider the expression of $\tilde{R}$ in modified gravity in eq. (\ref{eq5*}). To make $\tilde{R}<0$ we need  \begin{equation}
			\rho^{(e)}+3p^{(e)}<0,~~	|\rho^{(e)}+3p^{(e)}|> \rho+3p.\end{equation} Again it hints that $(\rho^{(e)},p^{(e)})$ corresponds to the density and pressure of dark energy if we assume that $(\rho,p)$ corresponds to the energy density and pressure of normal/usual matter that satisfies the SEC (i.e. $\rho+3p\geq0$). Thus we arrive at the same conclusion as in the former cases.
	\end{enumerate}
The above three points are consistent from cosmological point of view. The first point reveals that existence of singularity is not possible whenever there is accelerated expansion. This conclusion is supported in points 2 and 3. In point 2 we have Einstein gravity with two fluid system with one matter component that behaves as dark energy and it has been found that convergence of geodesic is possible only in the matter dominated era. In point 3 we have similar conclusion but using the modified gravity theory.\\ \\
	We shall now examine how Raychaudhuri scalar constraints the cosmological parameters. For flat FLRW model, if the modified gravity theory is equivalent to Einstein gravity with normal fluid (satisfying the SEC) and the effective matter is assumed to be in Dark energy form (not satisfying  Weak Energy Condition (WEC)) then it is possible to have some interrelation between the observationally measurable quantities and the Raychaudhuri scalar as follows: \\
	For normal matter considered as Dark matter with constant equation of state ($\omega$), the energy density from the conservation equation has the expression in terms of the redshift parameter $z$ as
	\begin{eqnarray}
		\rho=\rho_{0}(1+z)^{3(1+\omega)}.
	\end{eqnarray} where $\rho_{0}$ is the energy density at present.
	Now the first Friedmann equation
	\begin{eqnarray}
		H^{2}(z)=\dfrac{8 \pi G}{3}(\rho+\rho_{e})\label{eq15}
	\end{eqnarray} can be written in terms of the density parameter as
	\begin{equation}
		\tilde{H^{2}}=\Omega(1+z)^{3(1+\omega)} +(1-\Omega)\left(\frac{\rho_{e}}{\rho_{e_0}}\right).\label{eq16}
	\end{equation} Here, $\Omega=\dfrac{\rho_{0}}{\rho_{e}}$ is the density parameter for the dark matter, $\rho_{e}=\dfrac{3H_{0}^{2}}{8\pi G}$ is the critical density and $\tilde{H}=\dfrac{H}{H_{0}}$. Due to WEC for the effective fluid, $\dfrac{\rho_{e}}{\rho_{e_0}}\geq1$ for $z>0$, hence from equation (\ref{eq16}) $\tilde{H}$ has a lower bound
	\begin{eqnarray}
		H^{2}(z)\geq \Omega (1+z)^{3(1+\omega)}+(1-\Omega) \label{eq13}
	\end{eqnarray}
	As $\tilde{R}=q(\rho+\rho_{e})=q\rho_{e}\tilde{H^{2}}$, so the above inequality puts a restriction on $\tilde{R}$ as
	\begin{equation}
		\tilde{R}\geq q\rho_{e}\left[\Omega(1+z)^{3(1+\omega)} +(1-\Omega)\right]\label{eq18}
	\end{equation}
	Now, the luminosity distance $d_{L}$, a measurable quantity in the supernova red-shift survey is related to the coordinate distance $r(z)$ (defined as $	r(z)=\int_{0}^{z} \dfrac{dz'}{H(z')}$) by the relation:
	\begin{equation}
		d_{L}=c(1+z)\dfrac{r(z)}{H_{0}}.
	\end{equation}
	Thus the density parameter $\Omega$ and the above luminosity distance $d_{L}$ are restricted by the inequality (\ref{eq18}) as
	\begin{equation}
		\Omega\leq \left\{(1+z)^{3(1+\omega)}-1\right\}^{-1} \min \left[\dfrac{1}{(H_{0}r^{'}(z))^{2}},~\dfrac{\tilde{R}}{q \rho_{e}}\right]-1,
	\end{equation} and
	\begin{equation}
		d_{L}\leq \dfrac{c(1+z)}{H_{0}}\int_{0}^{z}\dfrac{dz'}{\left[\Omega(1+z')^{3(1+\omega)} +(1-\Omega)\right]^{\frac{1}{2}}}
	\end{equation}
	For a hyper-surface orthogonal congruence of time-like geodesic in isotropic background, the geometric form of the RE is given by
	\begin{equation}
		\dfrac{d\Theta}{d\tau}=-\dfrac{\Theta^{2}}{n}-\tilde{R}\label{eq23**}
	\end{equation}
	Using $\Theta=3H$ and equation (\ref{eq11}) the RE in terms of cosmic parameters can be written as 
	\begin{equation}
		\dot{H}=-(1+q)H^{2}\label{eq24**}
	\end{equation}
	Thus RE can be written in two ways: one in terms of geometric or kinematic variables namely equation (\ref{eq1}) or in particular form as equation (\ref{eq23**}) and the other in terms of cosmic parameters given by equation (\ref{eq24**}). We call the former as geometric form of RE and the later as cosmological form of RE. In this context, one may observe a nice analogy in these two forms. In geometric form $\tilde{R}$ or the Raychaudhuri scalar plays the role in convergence or focusing while in cosmological form $q$, the deceleration parameter does it. Moreover, the effect of gravity theory comes through $\tilde{R}$ in the geometric form while in the cosmological form $q$ carries the effect of gravity and this is clear from equation (\ref{eq8})-(\ref{eq9}). Therefore $q$ has a similar role as $\tilde{R}$. Further the geometric form of the RE can be converted to the evolution equation of a Harmonic Oscillator using a suitable transformation of variable. One may refer to \cite{Chakraborty:2023ork} for details of Harmonic oscillator equation derived from the geometric form of RE. The paper \cite{Chakraborty:2023ork} gives a transformation under which the first order RE can be converted to the Harmonic oscillator equation and it can be shown that the convergence condition, avoidance of singularity everything are related to the time varying frequency of the oscillator. Following this approach, in this paper we attempt to show the Harmonic oscillator equation from the cosmological form of the RE. For this, we consider the cosmological RE given by equation (\ref{eq24**}). Using,
	\begin{equation}
		H=\dfrac{\dot{Y}}{Y}
	\end{equation} the first order cosmological form of the RE can be converted to a second order differential equation analogous to the evolution equation of a classical real harmonic oscillator as 
	\begin{equation}
		\ddot{Y}+qH^{2}Y=0.\label{eq30}
	\end{equation} The frequency of the real harmonic oscillator is given by $W^{2}=qH^{2}$.  This harmonic oscillator equation is very much analogous to the Harmonic oscillator presented in \cite{Kar:2006ms} and \cite{Chakraborty:2023ork} where it was derived purely from the geometric form of the RE with suitable transformation of variables. Further the harmonic oscillator equation (\ref{eq30}) can be termed as cosmic Harmonic oscillator as it is expressed in terms of cosmological parameter $H$ and deceleration parameter $q$. Moreover, the above harmonic oscillator is a realistic one only in the matter dominated era and from \cite{Chakraborty:2023ork}, realistic harmonic oscillator is associated with the convergence of geodesics. Hence convergence of geodesic is only possible in the decelerated phase---a conclusion that we have already obtained earlier in this section.
	\section{Integrability of the Raychaudhuri equation and cosmological solutions}
	The Raychaudhuri equation for hyper-surface orthogonal congruence of time-like geodesics in FLRW space-time is given by equation (\ref{eq2}). In order to study the integrability of the RE we consider the following transformation
	\begin{equation}
		Z=\sqrt{h}=a^{3},
	\end{equation}
	where $h$=det($h_{\mu\nu}$) is the determinant of the metric of the n-dimensional space-like hyper-surface. The dynamical evolution of $h$ is given by,
	\begin{equation}
		\dfrac{1}{\sqrt{h}}\dfrac{d\sqrt{h}}{d\tau}=\Theta,
	\end{equation} so that \begin{equation}
		\dfrac{dZ}{d\tau}=Z\Theta.\end{equation}
	Hence for ($n$+1)-dimensional space-time manifold the RE can be written as a second order nonlinear ordinary differential equation as,
	\begin{equation}
		\dfrac{Z''}{Z}+ \left(\dfrac{1}{n}-1\right)\left(\dfrac{Z'}{Z}\right)^{2}+\tilde{R}=0.
	\end{equation} ` $'$ ' denotes differentiation w.r.t $\tau$. The above second order non-linear differential equation has a first integral of the form 
	\begin{equation}\label{eq22}
		H^{2}=\dfrac{a^{-\frac{6}{n}}}{9}\left[u_{0}-6\int a^{(\frac{6}{n}-1)}\tilde{R}~da\right],
	\end{equation} where $H=\dfrac{\dot{a}}{a}$ is the Hubble parameter, $a(t)$ is the scale factor and $u_{0}$ is a constant of integration.\\
	The above first integral for 4-D space-time ($n=3)$ can be identified as the first Friedmann equation
	\begin{equation} 3H^{2}+\dfrac{\kappa}{a^{2}}=\rho\label{eq36}
	\end{equation} with
	\begin{equation}
		\kappa=-\dfrac{u_{0}}{3}\label{eq23}
	\end{equation}
	and,
	\begin{equation}
		\rho=-\dfrac{2}{a^{2}}\int a~\tilde{R}~da\label{eq24}.
	\end{equation} Equation (\ref{eq23}) hints that $u_{0}$ is not merely a constant of integration but is related to the geometry of space-time as 
	$u_{0}>=<0$ for open/flat/closed model. Further one may show that (\ref{eq24}) holds in Einstein gravity as a particular case (see Appendix A).  In the above derivation we have obtained a first integral of the RE and it matches with the first Friedmann equation. Thus essentially, study of cosmology either by Einstein field equations or by RE seems to be identical. But RE seems to have an extra advantage. This is because it is a geometric theory, so it may hold not only in Einstein gravity but also in any other modified theories of gravity and it is reflected through equation (\ref{eq24}) where $\rho$ can be obtained using the geometric scalar $\tilde{R}$. It seems to be a general one as in the light of the current analysis the standard evolution is recovered (discussed later in this section). This may be treated as an advantage of this approach.
	\textbf{\underline{The cosmological solutions or the scenario of cosmic evolution using the above first integral (\ref{eq22})}}.\\ \\
	\textbf{CaseI :}   \underline{Matter in the form of perfect fluid with equation of state $p=\omega(a)\rho$.}\\
	In this case, $\tilde{R}=\dfrac{1}{2}\rho(1+3\omega(a))$. The energy-momentum conservation equation is 
	\begin{equation}
		\dot{\rho}+3\rho(1+\omega(a))H=0,
	\end{equation} the solution of which is given by
	\begin{equation}
		\rho=\rho_{0} a^{-3} \exp\left(-3\int \dfrac{w(a)}{a} da\right).
	\end{equation}
	So, 
	\begin{equation}
		\tilde{R}=\dfrac{\rho_{0}a^{-3}}{2}(1+3\omega(a))~ \exp\left(-3\int \dfrac{w(a)}{a}da\right).
	\end{equation}
	Hence from the first integral (\ref{eq22}) we have,
	\begin{equation}
		H^{2}=\dfrac{a^{-\frac{6}{n}}}{9}\left[u_{0}-3\rho_{0}\int a^{(\frac{6}{n}-4)}~(1+3\omega(a))~\exp\left(-3\int \dfrac{w(a)}{a} da\right)da\right].\label{eq42}
	\end{equation}
	\textbf{Subcase (i)}\underline{ $\omega(a)=0$ i.e dust era of evolution.}
	\begin{equation}
		H^{2}=\dfrac{u_{0}}{9} a^{\frac{-6}{n}}-\dfrac{\rho_{0}a^{-3}}{3(\frac{6}{n}-3)}.
	\end{equation} For 4D- space-time $n=3$, so
	\begin{equation}
		H^{2}=\dfrac{u_{0}}{9}a^{-2}+\dfrac{\rho_{0}}{3}a^{-3}.\label{eq44}
	\end{equation}
	Using $H=\dfrac{\dot{a}}{a}$ one has the solution as,
	\begin{equation}
		(t-t_{0})=3\int\dfrac{\sqrt{a}}{\sqrt{(3\rho_{0}+u_{0}a)}}da
	\end{equation}or,
	\begin{equation}
		(t-t_{0})=\dfrac{6}{u_{0}^{\frac{3}{2}}}\left[\dfrac{au_{0}}{2}\sqrt{(3\rho_{0}+au_{0})}-\dfrac{3\rho_{0}}{2}\cosh^{-1}\left(\sqrt{1+\frac{u_{0}a}{3\rho_{0}}}\right)+k\right],
	\end{equation} where $k$ is the constant of integration.
	Putting $u_{0}=0$ (flat space-time), for matter dominated era we have from equation (\ref{eq44})
		\begin{equation}
			H^{2}=\dfrac{\rho_{0}}{3}a^{-3},
		\end{equation} solving which we get the variation of scale factor $a(t)$ to cosmic time $t$ as $a(t)\propto t^{\frac{2}{3}}$, the standard result for matter dominated era.
	\begin{figure}[h!]
		\begin{minipage}{0.3\textwidth}
	~~~~~~~~~~~~~~~~~~~~~~~~~~~~~~~~~~~~~~~~~~~~~~~~~~~~~~~~		\centering\includegraphics[height=5cm,width=6cm]{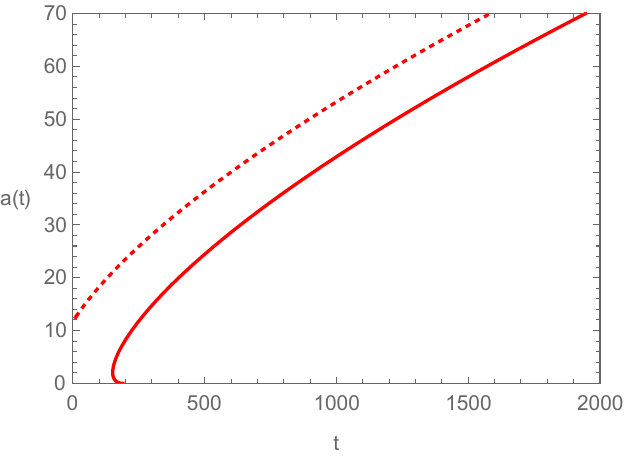}
		\end{minipage}
		\caption{[$a(t)$ vs $t$ for $\omega_0=0$ and (i) $u_0=0.1, k=1, \rho_0=1, t_0=0$ (Solid Red line); (ii) $u_0=0.01, k=-0.001, \rho_0=0.01, t_0=0$ ( Dotted Red line)]}\label{f1}
	\end{figure}
	\begin{figure}[h!]
		\begin{minipage}{0.3\textwidth}
	~~~~~~~~~~~~~~~~~~~~~~~~~~~~~~~~~~~~~~~~~~~~~~~~~~~~~~~~~~~		\centering\includegraphics[height=5cm,width=6cm]{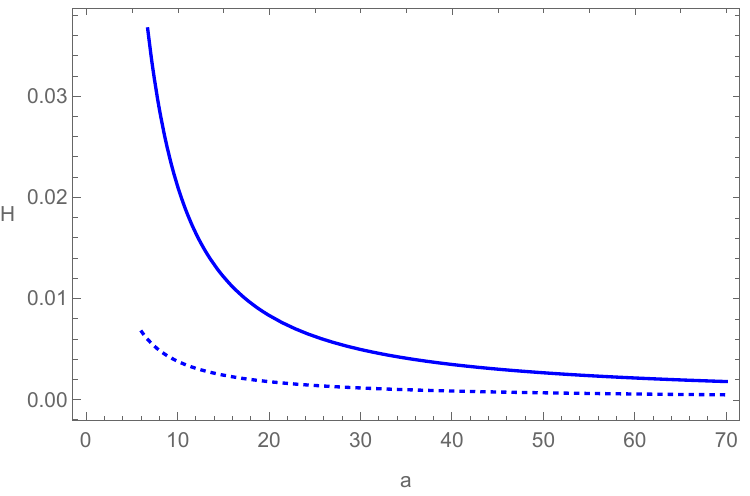}
		\end{minipage}
		\caption{[$H$ vs $a$ for $\omega_0=0$ and (i) $u_0=0.1, \rho_0=1$ (Solid Blue line); (ii) $u_0=0.01, \rho_0=0.01$ (Dotted Blue line)]}\label{f2}
	\end{figure}\\
	\textbf{Subcase (ii)} \underline{$\omega(a)=\omega_{0}$, a non-zero constant.}
	\begin{equation}
		H^{2}=\dfrac{a^{-\frac{6}{n}}}{9}\left[u_{0}-3\rho_{0}(1+3\omega_{0})\dfrac{a^{(\frac{6}{n}-3-3\omega_{0})}}{(\frac{6}{n}-3-3\omega_{0})}\right].
	\end{equation}
	For 4-D space-time again putting $n=3$, one gets
	\begin{equation}
		H^{2}=\dfrac{u_{0}}{9}a^{-2}+\dfrac{\rho_{0}}{3}a^{-3(1+\omega_{0})},
	\end{equation}or
	\begin{equation}
		(t-t_{0})=3\int\dfrac{da}{\sqrt{u_{0}+3\rho_{0}a^{-(1+3\omega_{0})}}}.
	\end{equation}
For the radiation dominated era characterized by $\omega_{0}=\dfrac{1}{3}$, using equation (\ref{eq42}) we have,
		\begin{equation}
			H^{2}=3\rho_{0}a^{-4}
		\end{equation} which gives the variation of scale factor to cosmic time as $a(t)\propto t^{\frac{1}{2}}$. To find the vacuum dominated solution we put $\kappa=-\dfrac{u_{0}}{3}=0$ and $\rho=\rho_{0}$, a constant in equation (\ref{eq36}). Then we get
		\begin{equation}
			3H^{2}=\rho_{0}
		\end{equation} so that the variation of scale factor to cosmic time is $a(t)\propto \exp(\Lambda t)$ where $\Lambda=\sqrt{\dfrac{\rho_{0}}{3}}$.
	\begin{table}
		\begin{tabular}{ |p{2cm}|p{9cm}|}
			\hline
			\multicolumn{2}{|c|}{Solution for various non-zero choices of $\omega_{0}$} \\
			\hline
			$\omega_{0}$& Solution  \\
			\hline
			$~~\frac{1}{3}$  &  $a(t)=\left[\dfrac{(t-t_{0})^{2}u_{0}}{9}-\dfrac{3\rho_{0}}{u_{0}}\right]^{\frac{1}{2}}$   \\
			$~~1$& $(t-t_{0})=\dfrac{a^{3}~2^{F_{1}}~\left(\frac{1}{2},\frac{3}{4};\frac{7}{4};\frac{-u_{0}}{3\rho_{0}}a^{4}\right)}{\sqrt{3\rho_{0}}}$  \\
			$-\frac{1}{3}$ & 
			$a(t)=\dfrac{\sqrt{u_{0}+3\rho_{0}}}{3}(t-t_{0})$ \\
			$-1$ & $a(t)=\sqrt{\frac{u_{0}}{3\rho_{0}}}\left[\coth^{2}(\sqrt{\frac{\rho_{0}}{3}}(t-t_{0}))-1\right]^{\frac{-1}{2}}$\\
			\hline
		\end{tabular}
		\caption{Cosmic scale factor for various non zero choices of $\omega=\omega_{0}$}\label{T01}
	\end{table} Thus, in light of the current analysis how the standard evolution is recovered has been shown. 	However, the Raychaudhuri based general solutions for various non-zero choices (of course that make the integration solvable) of $\omega_{0}$ are given below in TABLE \ref{T01}.
	\begin{figure}[h!]
		\begin{minipage}{0.3\textwidth}
~~~~~~~~~~~~~~~~~~~~~~~~~~~~~~~~~~~~~~~~~~~~~~~~~~~~~~~~~~~~			\centering\includegraphics[height=5cm,width=6cm]{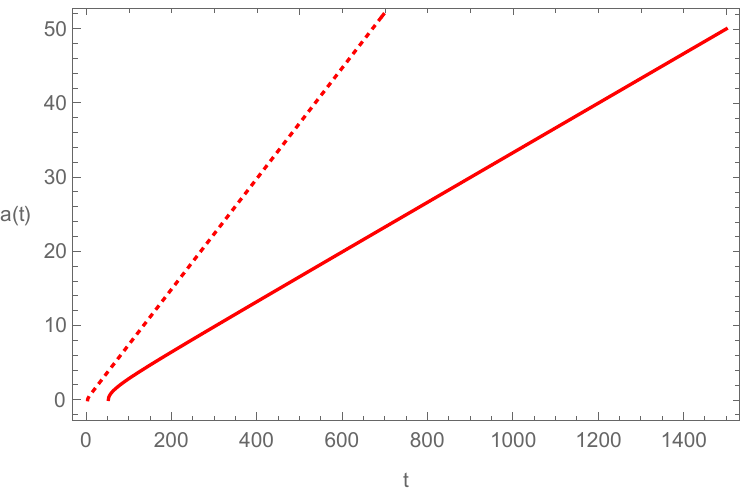}
		\end{minipage}
		\caption{[$a(t)$ vs $t$ for $\omega_0=\dfrac{1}{3}$ and (i) $u_0=0.01, \rho_0=0.01, t_0=0$ (Solid Red line); (ii) $u_0=0.05, \rho_0=0.001, t_0=0$ ( Dotted Red line)]}\label{f3}
	\end{figure}\\
	\begin{figure}[h!]
		\begin{minipage}{0.3\textwidth}
~~~~~~~~~~~~~~~~~~~~~~~~~~~~~~~~~~~~~~~~~~~~~~~~~~~~~~~~~~~~~			\centering\includegraphics[height=5cm,width=6cm]{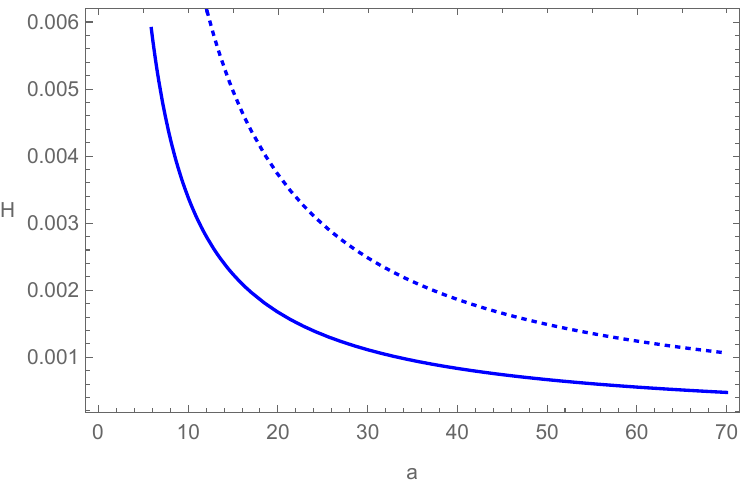}
		\end{minipage}
		\caption{[$H$ vs $a$ for $\omega_0=\dfrac{1}{3}$ and (i) $u_0=0.01, \rho_0=0.01$ (Solid Blue line); (ii) $u_0=0.05, \rho_0=0.001$ (Dotted Blue line)]}\label{f4}
	\end{figure}\\
	\begin{figure}[h!]
		\begin{minipage}{0.3\textwidth}
~~~~~~~~~~~~~~~~~~~~~~~~~~~~~~~~~~~~~~~~~~~~~~~~~~~~~~~~~~~~~~~		\centering\includegraphics[height=5cm,width=6cm]{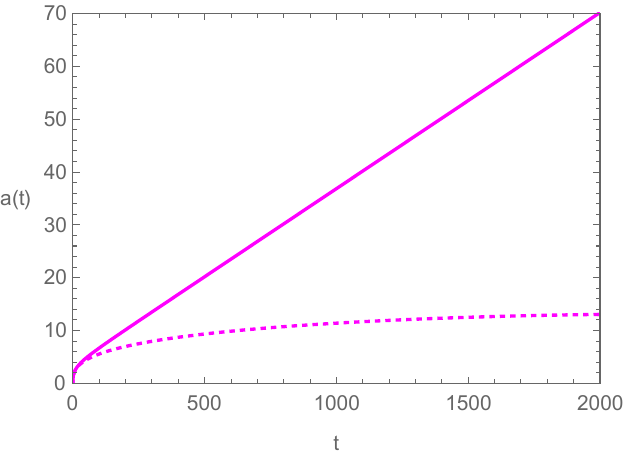}
		\end{minipage}
		\caption{[$a(t)$ vs $t$ for $\omega_0=1$ and (i) $u_0=0.01, \rho_0=1, k=0.1, t_0=0$ (Solid Magenta line); (ii) $u_0=-0.0001, \rho_0=1, k=0.1, t_0=0$ ( Dotted Magenta line)]}\label{f5}
	\end{figure}
	\begin{figure}[h!]
		\begin{minipage}{0.3\textwidth}
~~~~~~~~~~~~~~~~~~~~~~~~~~~~~~~~~~~~~~~~~~~~~~~~~~~~~~~~~~~~~~~~		\centering\includegraphics[height=5cm,width=6cm]{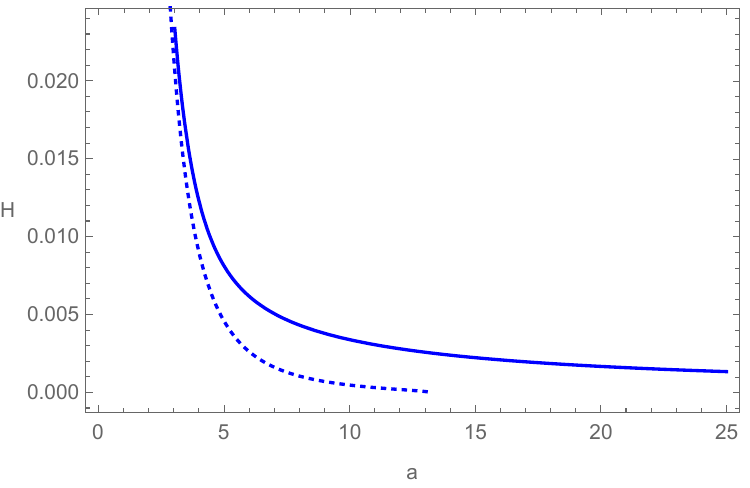}
		\end{minipage}
		\caption{[$H$ vs $a$ for $\omega_0=1$ and (i) $u_0=0.01, \rho_0=1$ (Solid Blue line); (ii) $u_0=-0.0001, \rho_0=1$ ( Dotted Blue line)]}\label{f6}
	\end{figure}
	\begin{figure}[h!]
		\begin{minipage}{0.3\textwidth}
			\centering\includegraphics[height=5cm,width=6cm]{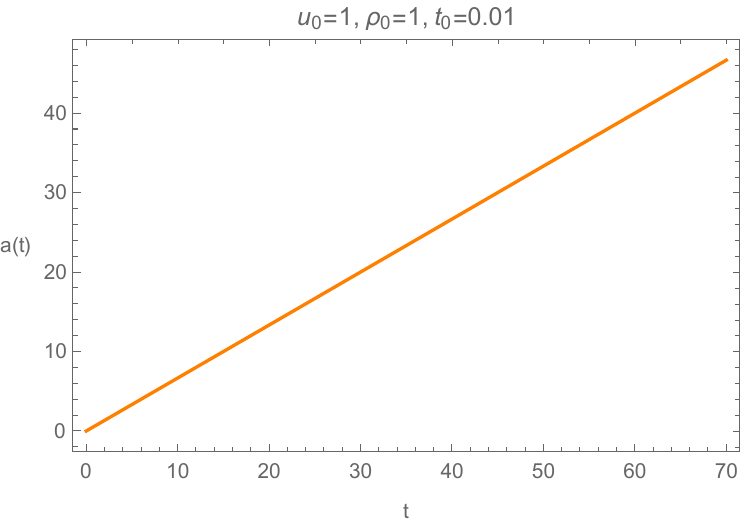}
		\end{minipage}~~~~~~~~~~~~
		\begin{minipage}{0.3\textwidth}
			\centering\includegraphics[height=5cm,width=6cm]{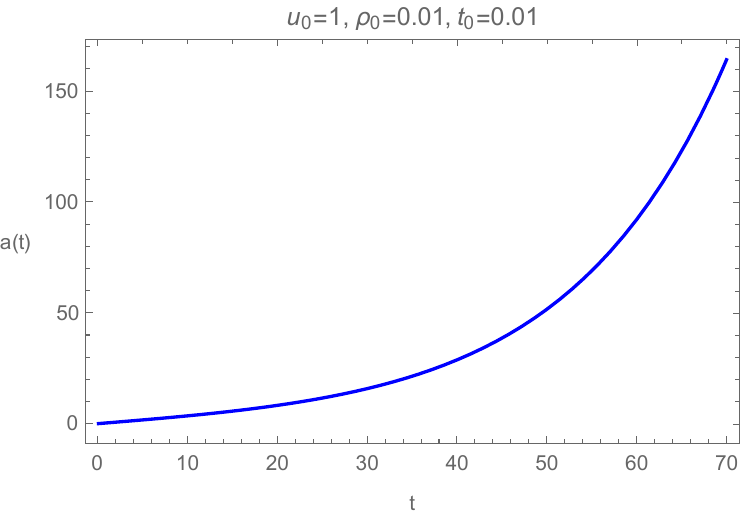}\end{minipage}
		\begin{minipage}{0.85\textwidth}
			\caption{[$a(t)$ vs $t$ for $\omega_0=-\dfrac{1}{3}$ (left) and for $\omega_0=-1$ (right)]}\label{f7}
		\end{minipage}
	\end{figure}
	\begin{figure}[h!]
		\begin{minipage}{0.3\textwidth}
			\centering\includegraphics[height=5cm,width=6cm]{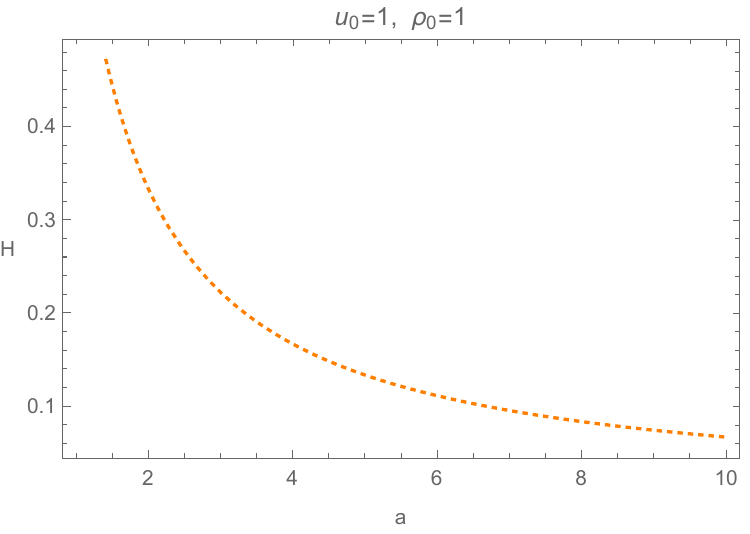}
		\end{minipage}~~~~~~~~~~~~
		\begin{minipage}{0.3\textwidth}
			\centering\includegraphics[height=5cm,width=6cm]{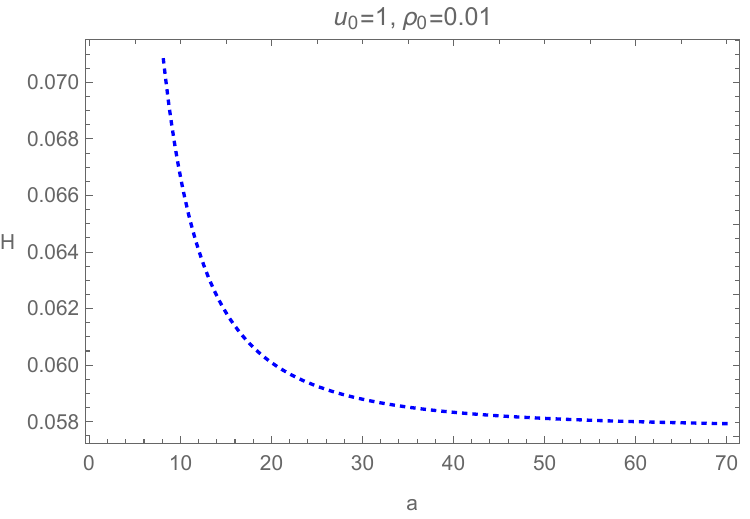}\end{minipage}
		\begin{minipage}{0.85\textwidth}
			\caption{[$H$ vs $a$ for $\omega_0=-\dfrac{1}{3}$ (left) and for $\omega_0=-1$ (right)]}\label{f8}
		\end{minipage}
	\end{figure}
\newpage
	The solution for a general $w(a)$ of the form [for ref. see  	\cite{Jassal:2005qc}] \begin{equation} 
		\omega(a)=\omega_{0}+\omega^{'}\left(\dfrac{a}{a-1}\right),\end{equation}is given by
	\begin{equation}
		\dot{a}=\dfrac{1}{3}\left[u_{0}-3\rho_{0}\int((1+3\omega_{0})(a-1)+3\omega^{'}a)(a-1)^{-1-3\omega^{'}}a^{-(2+3\omega_{0})}da\right]^{\frac{1}{2}}
	\end{equation} which upon further simplification yields
	\begin{equation}
		3\int\dfrac{da}{\left[u_{0}-3\rho_{0}~A\right]^{\frac{1}{2}}}=(t-t_{0}),
	\end{equation} where $A$ is given by
	\begin{equation}
		\scriptsize	A=\dfrac{(a-1)^{-3\omega^{'}}}{3\omega^{'}}\left[-(1+3\omega_{0}+3\omega^{'})~2^{F_{1}}(1+3\omega_{0},-3\omega^{'};1-3\omega^{'};1-a)+(1+3\omega_{0})~2^{F_{1}}~(2+3\omega_{0},-3\omega^{'};1-3\omega^{'};1-a)\right].
	\end{equation}
	and $2^{F_{1}}$ is the Gauss-Hypergeometric function.
	\begin{figure}[h!]
		\begin{minipage}{0.3\textwidth}
			\centering\includegraphics[height=5cm,width=6cm]{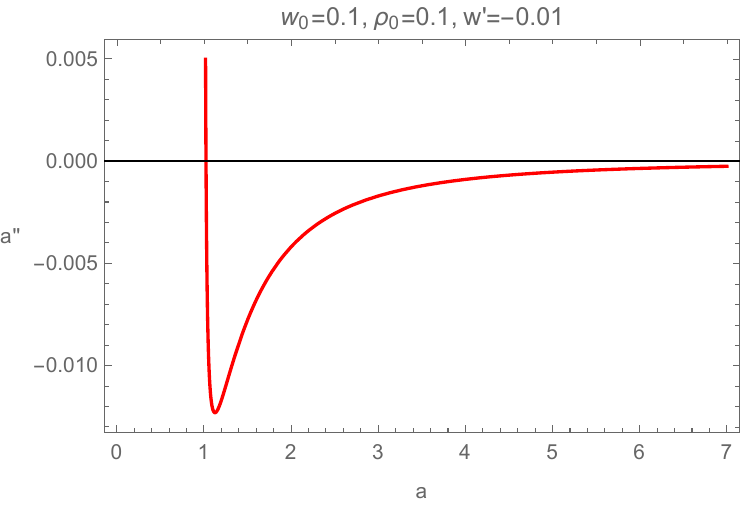}
		\end{minipage}~~~~~~~~~~~~
		\begin{minipage}{0.3\textwidth}
			\centering\includegraphics[height=5cm,width=6cm]{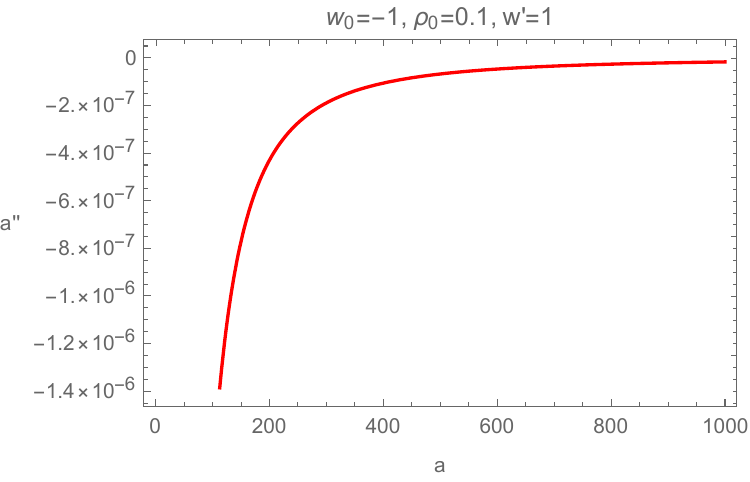}
		\end{minipage}
		\begin{minipage}{0.3\textwidth}
			\centering\includegraphics[height=5cm,width=6cm]{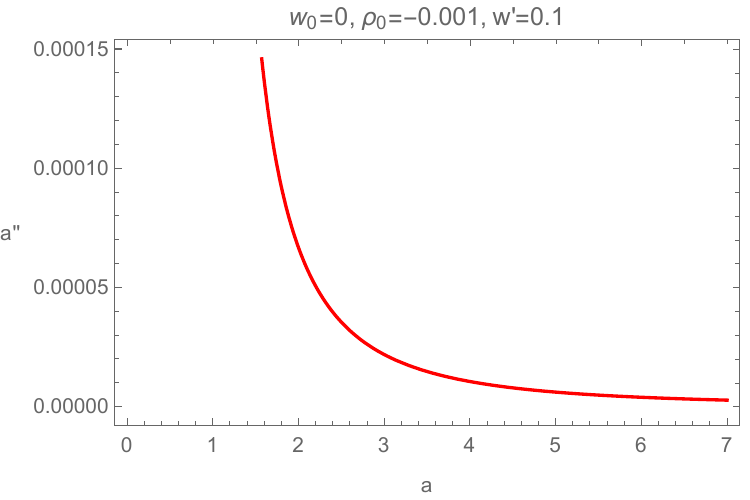}
		\end{minipage}~~~~~~~~~~~
		\begin{minipage}{0.3\textwidth}
	~~~~~~~~~~~~~~~~~~~~~~~~~~~~~~~~~~~~~~~~~~~~~~~~~~~~~~~~~~~~~~~		\centering\includegraphics[height=5cm,width=6cm]{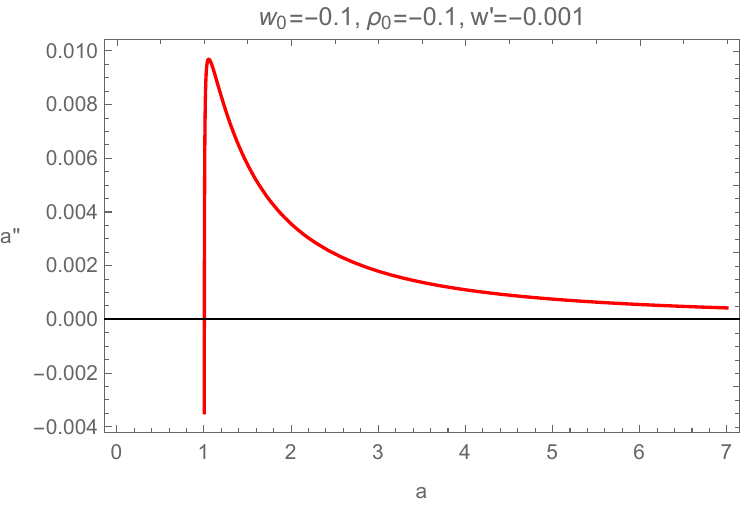}
		\end{minipage}
		\caption{[ $a''$ vs $a$ for $\omega(a)=\omega_{0}+\omega^{'}\left(\dfrac{a}{a-1}\right)$ and various choices of the parameters specified in each panel]}\label{f9}
	\end{figure}
	\newpage
	\textbf{Case II-}  \underline{Inflation:}\\
	In case of inflation $\tilde{R}=\dfrac{1}{2}(\rho_{\phi}+3p_{\phi})\simeq-V_{0}$. So one has
	\begin{equation}
		H^{2}=\dfrac{nV_{0}}{9}+\dfrac{u_{0}}{9}a^{-\frac{6}{n}}.
	\end{equation}
	For 4-D space-time the solution becomes,
	\begin{equation}
		(t-t_{0})=\int\dfrac{da}{\sqrt{\frac{u_{0}}{9}+\frac{V_{0}a^{2}}{3}}},
	\end{equation}or
	\begin{equation}
		(t-t_{0})=\sqrt{\dfrac{3}{V_{0}}}~\ln|a|+\left(a^{2}+\frac{u_{0}}{3V_{0}}\right)^{\frac{1}{2}}+C,
	\end{equation} $C$, being the constant of integration.\\
	\begin{figure}[h!]
		\begin{minipage}{0.3\textwidth}
~~~~~~~~~~~~~~~~~~~~~~~~~~~~~~~~~~~~~~~~~~~~~~~~~~~~~~~~~~~~~			\centering\includegraphics[height=5cm,width=6cm]{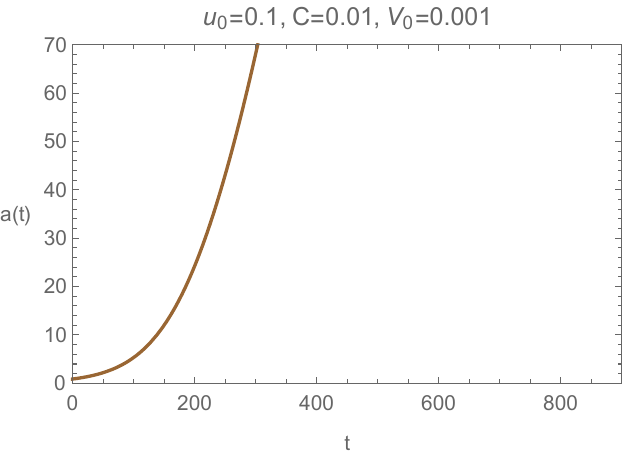}
		\end{minipage}
		\caption{[Graphical representation of scale factor $a(t)$ with cosmic time $t$ during inflation]}\label{f10}
	\end{figure}
	\section{Brief discussion and conclusion}
	The present work is an example where the RE has been used in cosmological context. A general formulation of the Raychaudhuri scalar (i.e.Curvature scalar) has been done in terms of cosmological parameters both in Einstein gravity as well as in modified gravity theories. It is found that the positiveness of the curvature scalar or precisely the convergence condition demands that $q$, the deceleration parameter should be positive. Hence, the deceleration parameter for the present universe being negative does not allow focusing theorem to hold. Again since $q=-1(<0)$ during the inflationary era of evolution, therefore one may conclude that matter dominated era of evolution is in favor of convergence while the formation singularity can be avoided both in early inflationary era and in the present accelerated era of expansion. Further, a nice analogy of RE through geometry and cosmology has been elaborated and the role of $\tilde{R}$ and $q$ has been shown to be same in the context of focusing. Using a suitable transformation the cosmological form of RE has been converted a Harmonic oscillator equation and condition for the formation of real Harmonic oscillator has been shown to be possible only in the matter dominated era of cosmic evolution. The observable quantities namely luminosity distance ($d_L$) and density parameter ($\Omega$) are shown to be related to the curvature scalar ($\tilde{R}$). Using a suitable transformation related to geometric variable (metric scalar of the hyper-surface), the RE is transformed to a second order non-linear differential equation whose first integral is easily obtained. It is found that this first integral is nothing but the 1st Friedmann equation. By choosing perfect fluid with barotropic equation of state as the matter content of the universe, cosmological solutions are obtained for various choices of the parameters in the equation of state. In most of the choices (where scale factor and Hubble parameter have explicit form), the cosmological parameters namely the scale factor ($a$) and the Hubble parameter ($H$) behave in accordance with the observational data at least qualitatively i.e. the universe is in an expanding phase with the rate of expansion gradually decreases.  However, for the specific choices of the equation of state as a function of the scale factor (variable equation of state) only acceleration ($\ddot{a}$) can be evaluated as a function of the scale factor ($a$) in FIG.s (\ref{f9}). Due to some choices of the parameters involved it is found that: (i) there is a sharp fall from acceleration to deceleration and asymptotically it goes to zero acceleration (FIG. \ref{f9} top left), (ii) the universe is totally in a decelerated phase with rate of decrease of deceleration being sharp in the initial stage and then gradually becomes a constant (FIG. \ref{f9} top right), (iii) the universe experiences only accelerating phase but it gradually decreases with the evolution and finally reaches a constant value (FIG. \ref{f9} bottom left), (iv) there is a sharp rise from deceleration to acceleration and then gradually the acceleration parameter goes to zero asymptotically (FIG. \ref{f9} bottom right).
	
 Finally it is to be noted that throughout the paper we have used congruence of time-like geodesics not the congruence of null geodesic. In principle, there is no basic difference between the use of these two kinds of geodesics. However, in the context of cosmology as the evolution is characterized by time variation so time-like geodesic is important where proper time or the cosmic time is being used. In case of null geodesics since we have only an affine parameter, so time evolution will be rather ambiguous. One may interpret an affine parameter to be the null analog of proper time. It is worthy to mention that there is a nice study in the literature \cite{Albareti:2012se} where the convergence scalar has been geometrically interpreted as the mean curvature. Using this interpretation, in the present context we can say that the frequency of the harmonic oscillator formed out of the cosmological RE is associated to the mean curvature. It would be interesting to find the solution of this harmonic oscillator equation subject to some realistic choices of the mean curvature in future work.  Also the authors in \cite{Albareti:2012se}, have analyzed cosmic evolution from the expression of deceleration parameter $q$ based on $\Lambda CDM$ model. On the other hand, this paper shows the complete evolution of the universe by analyzing the first integral of the Raychaudhuri equation.
	\section*{ACKNOWLEDGMENT}
The authors thank the anonymous reviewers for their valuable and insightful comments, questions and suggestions which increased the quality and visibility of the work. The author M.C thanks University Grant Commission (UGC) for providing the Junior Research Fellowship (ID:211610035684/JOINT CSIR-UGC NET JUNE-2021). S.C. thanks FIST program of DST, Department of Mathematics, JU (SR/FST/MS-II/2021/101(C)).\\
	\section*{\underline{Appendix A}}
	To show the equivalence between the first integral of the RE and the first Friedmann equation in FLRW model, we consider the Einstein field equations,
	\begin{equation}
		G_{\mu\nu}=R_{\mu\nu}-\frac{1}{2}~R~g_{\mu\nu}=T_{\mu\nu},
	\end{equation} or
	\begin{eqnarray}
		R_{\mu\nu}= T_{\mu\nu}-\dfrac{1}{2}Tg_{\mu\nu}\\
		\therefore	\tilde{R}= R_{\mu\nu}u^{\mu}u^{\nu}=T_{\mu\nu}u^{\mu}u^{\nu}+\dfrac{1}{2}T.\\ ~(T=g^{\mu\nu}T_{\mu\nu})\nonumber
	\end{eqnarray}
	If the matter content is assumed to be perfect fluid then the energy-momentum tensor is given by
	\begin{equation}
		T_{\mu\nu}=pg_{\mu\nu}+(\rho+p)u^{\mu}u^{\nu}.
	\end{equation} From the solution of energy-conservation equation $\dot{\rho}+3H(p+\rho)=0$ one has
	\begin{equation}
		\rho=\rho_{0}a^{-3(1+\omega)},
	\end{equation} where $\omega$ is the equation of state parameter ($p=\omega \rho$).  Hence $\tilde{R}=\dfrac{\rho+3p}{2}=\dfrac{(1+3\omega)}{2}\rho_{0}a^{-3(1+\omega)}$. Substituting this $\tilde{R}$ in the R.H.S of equation (\ref{eq24}) one gets $\rho$, the L.H.S of equation (\ref{eq24}) i.e.\\
	$\frac{-2}{a^{2}}\int a \tilde{R}~da=\frac{-2}{a^{2}}\frac{(1+3\omega)\rho_{0}}{2}\int a^{-(2+3\omega)~da}=\frac{-2}{a^{2}}\times\frac{-\rho_{0}}{2}\times a^{-(1+3\omega)}=\rho_{0}a^{-3(1+\omega)}=\rho.$
	Thus (\ref{eq24}) holds good in Einstein gravity for a 4-D space-time as a particular case.
	\section*{\underline{Appendix B}}
	\subsection*{Singularity and Geodesic incompleteness: A general study by Raychaudhuri equation}
 A space-time manifold is causal geodesically incomplete if any one of the following holds:
		\begin{itemize}
			\item a proper condition on the curvature i.e, a proper energy condition.
			\item an appropriate condition on the causality.
			\item a suitable initial or boundary condition.
		\end{itemize}
		The initial condition serves the purpose that some causal geodesics start to focus towards each other then the energy condition implies that this focusing goes on till a causal geodesic develops a focal/conjugate point. As a result, the geodesic stops maximizing the Lorentzian distance. The causality condition on the other hand implies the existence of maximizing geodesic atleast in some region of space-time. Thus, to resolve the above contradiction it is expected that the geodesic should terminate before they reach a conjugate/focal point i.e, they are incomplete in nature. From the point of view of the singularity theorems the above curvature condition is necessary for focusing effect on causal geodesics. Mathematically, the central idea is to introduce the Jacobi field, a vector field along the geodesic $\gamma$ which satisfies the Jacobi equation
		\begin{equation}
			\ddot{J}+R(J, \dot{\gamma})\dot{\gamma}=0.\nonumber
		\end{equation} One can interpret this Jacobi field as a one-one correspondence with geodesic variation of $\gamma$ to have a clear picture.
		
		For focusing of geodesics, the notion of conjugate points is essential. Points $\gamma(a)$ and $\gamma(b)$ on the geodesic are called conjugate if there exists a non-trivial Jacobi field which vanishes at $a$ and $b$. It has been established that a causal geodesic fails to maximize the Lorentzian distance after its first conjugate point. To have an analytic tool for determination of conjugate points, one may note that the relevant information on the conjugate points is contained in the $(n-1)$-dimensional subspace of the Jacobi field vanishing at a given point and taking values in the set $\dot{\gamma(t)}^{\perp}:=\{v\in T_{\gamma(t)}(M):<v,\dot{\gamma(t)}>=0\}$.
		
		Let us now define a class of $(1,1)$ tensor field matrix $[A]:[\dot{\gamma}]^{\perp}\rightarrow[\dot{\gamma}]^{\perp}$ for which the tensor Jacobi equation
		\begin{equation}
			[\ddot{A}]+[R][A]=0.\nonumber
		\end{equation}
		Here $R$ can be treated as a tidal force operator i.e, $[R]:[v]\rightarrow [R(v,\dot{\gamma})\dot{\gamma}]$. Now the analytic way to detect conjugate point can be obtained through Raychaudhuri equation 
		\begin{equation}
			\dot{\Theta}=-Ric(\dot{\gamma},\dot{\gamma})-tr(\Sigma^{2})-\dfrac{\Theta^{2}}{D}\nonumber
		\end{equation} where the expansion scalar $\Theta$ is given by $\Theta=tr([\dot{A}][A]^{-1})=(det[A])^{-1}(det[A])^{^.}$ and shear $\Sigma$ is defined as $\Sigma=\dfrac{1}{2}([B]+[B^{\dag}])-\dfrac{\Theta}{D}i$d with $B=[\dot{A}][A^{-1}]$. In this RE, the second and third term on the right hand side are negative definite. If the first term is also non positive through the SEC or NEC then one can generate conjugate points. In particular $\Theta(a)<0$ at some parameter value $a$ then  it will diverge to $-\infty$ in finite parameter time. Further from the above RE we have $\ddot{\Theta}\leq-\dfrac{\Theta^{2}}{D}$ which on integration from $a$ to some $t>a$ gives 
		\begin{equation}
			\Theta\leq\dfrac{D}{t-a+\frac{D}{\Theta(a)}}\nonumber
		\end{equation}
		Thus, $\Theta$ diverges for some interval $a\leq t< a-\frac{D}{\Theta(a)}$. Thus if $[A]$ is a Jacobi tensor class with $[A](a)=0$ and $\dot{[A]}(a)$=id, the identity mapping and we have $|\Theta(t)|\rightarrow \infty$ for $t$ to some $b$ then $\det[A(b)]=0$ and hence $\gamma_{b}$ is conjugate to $\gamma_{a}$.  Hence the causal geodesic $\gamma$ becomes inextendible after the first conjugate point. In this way the notion of incomplete geodesic, existence of conjugate point and divergence of $\Theta$ can be associated using the geometrical RE.
	

\begin{thebibliography}{50}
		\bibitem{SupernovaSearchTeam:1998fmf}
		A.~G.~Riess \textit{et al.} [Supernova Search Team],
		Astron. J. \textbf{116}, 1009-1038 (1998)
		\bibitem{SupernovaCosmologyProject:1998vns}
		S.~Perlmutter \textit{et al.} [Supernova Cosmology Project],
		Astrophys. J. \textbf{517}, 565-586 (1999)
		\bibitem{WMAP:2003elm}
		D.~N.~Spergel \textit{et al.} [WMAP],
		Astrophys. J. Suppl. \textbf{148}, 175-194 (2003)
		\bibitem{SDSS:2003eyi}
		M.~Tegmark \textit{et al.} [SDSS],
		Phys. Rev. D \textbf{69}, 103501 (2004)
		\bibitem{SDSS:2005xqv}
		D.~J.~Eisenstein \textit{et al.} [SDSS],
		Astrophys. J. \textbf{633}, 560-574 (2005)
		\bibitem{Poisson:2009pwt}
		E.~Poisson,
		``A Relativist's Toolkit: The Mathematics of Black-Hole Mechanics,
		Cambridge University Press, 2009,
		doi:10.1017/CBO9780511606601
		\bibitem{Raychaudhuri:1953yv}
		A.~Raychaudhuri,
		Phys. Rev. \textbf{98}, 1123-1126 (1955)
		\bibitem{raychaudhuri}
		Raychaudhuri, Amal K. "Theoretical cosmology." Clarendon  Press , Oxford Studies in Physics (1979)
		\bibitem{LIGOScientific:2017vwq}
		B.~P.~Abbott \textit{et al.} [LIGO Scientific and Virgo],
		Phys. Rev. Lett. \textbf{119}, no.16, 161101 (2017)
		\bibitem{Senovilla:2014gza}
			J.~M.~M.~Senovilla and D.~Garfinkle,
			Class. Quant. Grav. \textbf{32}, no.12, 124008 (2015)
		\bibitem{Hawking:1970zqf}
		S.~W.~Hawking and R.~Penrose,
		Proc. Roy. Soc. Lond. A \textbf{314}, 529-548 (1970)
		\bibitem{Hawking:1973uf}
		S.~W.~Hawking and G.~F.~R.~Ellis,
		``The Large Scale Structure of Space-Time,''
		Cambridge University Press, 2023,
		doi:10.1017/9781009253161
		\bibitem{Penrose:1964wq}
		R.~Penrose,
		Phys. Rev. Lett. \textbf{14}, 57-59 (1965)
		\bibitem{Steinbauer:2022hvq}
			R.~Steinbauer,
			``The singularity theorems of General Relativity and their low regularity extensions,''
			[arXiv:2206.05939 [math-ph]].
		\bibitem{Dadhich:2007pi}
		N.~Dadhich,
		Pramana \textbf{69}, 23-30 (2007)
		\bibitem{Burger:2018hpz}
		D.~J.~Burger, N.~Moynihan, S.~Das, S.~Shajidul Haque and B.~Underwood,
		Phys. Rev. D \textbf{98}, no.2, 024006 (2018)
		
		\bibitem{Kar:2006ms}
		S.~Kar and S.~SenGupta,
		Pramana \textbf{69}, 49 (2007)
		\bibitem{Ehlers:2006aa}
		J.~Ehlers,
		Int. J. Mod. Phys. D \textbf{15}, 1573-1580 (2006)
		\bibitem{Kar:2008zz}
		S.~Kar,
		Resonance J. Sci. Educ. \textbf{13}, 319-333 (2008)
		
		\bibitem{Horwitz:2021lyc}
		L.~P.~Horwitz, V.~S.~Namboothiri, G.~Varma K, A.~Yahalom, Y.~Strauss and J.~Levitan,
		Symmetry \textbf{13}, no.6, 957 (2021)
		
		\bibitem{Dadhich:2005qr}
		N.~Dadhich,
		``Derivation of the Raychaudhuri equation,''18 Nov (2022) [arXiv:gr-qc/0511123 [gr-qc]]
		\bibitem{Chakraborty:2023ork}
		M.~Chakraborty, A.~Bose and S.~Chakraborty,
		Phys. Scripta \textbf{98}, no.2, 025007 (2023)
		\bibitem{Chakraborty:2023yyz}
		M.~Chakraborty and S.~Chakraborty,
		Class. Quant. Grav. \textbf{40}, no.15, 155010 (2023)
		\bibitem{Choudhury:2021zij}
		S.~G.~Choudhury, A.~Dasgupta and N.~Banerjee,
		Int. J. Geom. Meth. Mod. Phys. \textbf{18}, no.08, 2150115 (2021)
		\bibitem{Bhatt:2021hdi}
		R.~P.~Bhatt, A.~Roy and S.~Kar,
		Reson. \textbf{28}, no.3, 389-410 (2023)
		\bibitem{Dasgupta:2007nr}
		A.~Dasgupta, H.~Nandan and S.~Kar,
		Annals Phys. \textbf{323}, 1621-1643 (2008)
		\bibitem{Chakraborty:2023rgb}
		M.~Chakraborty and S.~Chakraborty,
		Annals Phys. \textbf{460}, 169577 (2024)
		\bibitem{Chakraborty:2023voy}
		M.~Chakraborty and S.~Chakraborty,
		Annals Phys. \textbf{457}, 169403 (2023)
		\bibitem{Das:2013oda}
		S.~Das,
		Phys. Rev. D \textbf{89}, no.8, 084068 (2014)
		\bibitem{Chakraborty:2023lav}
		M.~Chakraborty and S.~Chakraborty,
		Mod. Phys. Lett. A \textbf{38}, no.28n29, 2350129 (2023)
		\bibitem{Blanchette:2020kkk}
		K.~Blanchette, S.~Das, S.~Hergott and S.~Rastgoo,
		Phys. Rev. D \textbf{103}, no.8, 084038 (2021)
		\bibitem{Albareti:2014dxa}
			F.~D.~Albareti, J.~A.~R.~Cembranos, A.~de la Cruz-Dombriz and A.~Dobado,
			JCAP \textbf{03}, 012 (2014)
		\bibitem{Albareti:2012se}
			F.~D.~Albareti, J.~A.~R.~Cembranos and A.~de la Cruz-Dombriz,
			JCAP \textbf{12}, 020 (2012)
		\bibitem{Jassal:2005qc}
		H.~K.~Jassal, J.~S.~Bagla and T.~Padmanabhan,
		Phys. Rev. D \textbf{72}, 103503 (2005)
	\end{thebibliography}
\end{document}